\documentclass[conference]{IEEEtran}
\IEEEoverridecommandlockouts
\usepackage{cite}
\usepackage{amsmath,amssymb,amsfonts,pgfplots}
\usepackage{algorithmic}
\usepackage{graphicx}
\usepackage{textcomp}
\usepackage{xcolor}
\usepackage{bm}

\usepackage{booktabs}
\usepackage{makecell}

\def\BibTeX{{\rm B\kern-.05em{\sc i\kern-.025em b}\kern-.08em
    T\kern-.1667em\lower.7ex\hbox{E}\kern-.125emX}}

\begin{document}

\title{MUKA: MULTI KERNEL AUDIO ADAPTATION OF AUDIO-LANGUAGE MODELS}

\author{\IEEEauthorblockN{Reda Bensaid}
\IEEEauthorblockA{\textit{IMT Atlantique, Brest, France} \\
\textit{Polytechnique Montréal, Canada} \\
reda.bensaid@imt-atlantique.fr}
\and
\IEEEauthorblockN{Amine Ouasfi}
\IEEEauthorblockA{\textit{Inria, University Rennes, IRISA, CNRS} \\
amine.ouasfi@inria.fr}
\and
\IEEEauthorblockN{Yassir Bendou}
\IEEEauthorblockA{\textit{IMT Atlantique, Brest, France} \\
yassir.bendou@gmail.com}
\and
\IEEEauthorblockN{Ilyass Moummad}
\IEEEauthorblockA{\textit{Inria, LIRMM, Université de Montpellier} \\
ilyass.moummad@inria.fr}
\and
\IEEEauthorblockN{Vincent Gripon}
\IEEEauthorblockA{\textit{IMT Atlantique, Brest, France} \\
vincent.gripon@imt-atlantique.fr}
\and

\IEEEauthorblockN{François Leduc-Primeau}
\IEEEauthorblockA{\textit{Polytechnique Montréal, Canada} \\
francois.leduc-primeau@polymtl.ca}

\and

\IEEEauthorblockN{Adnane Boukhayma}
\IEEEauthorblockA{\textit{Inria, University Rennes, IRISA, CNRS} \\
adnane.boukhayma@inria.fr}
}
\maketitle

\begin{abstract}
Multimodal foundation models have demonstrated impressive generalization capabilities, yet efficiently adapting them to new tasks in a few-shot setting remains a critical challenge. In this work, we investigate the few-shot adaptation of Large Audio-Language Models (ALMs) through both training-based and training-free approaches. We introduce MUKA, a multi-kernel adaptation framework that combines the fine-grained, context-dependent representations of instruction-tuning based models like Pengi with the global semantic representations of  contrastive pretraining methods like CLAP. By constructing a product kernel that aligns local similarity with global semantics, MUKA enhances representational power while preserving the theoretical guarantees of kernel methods and avoiding additional training. Extensive experiments across 11 diverse audio datasets demonstrate that MUKA achieves state-of-the-art performance among training-free methods and even surpasses training-based adapters in several scenarios, offering a compelling balance between adaptability and efficiency. 
\end{abstract}


\section{Introduction}
\label{sec:intro}

\begin{figure}[!t]
\centering
\centerline{\includegraphics[width=0.5\textwidth]{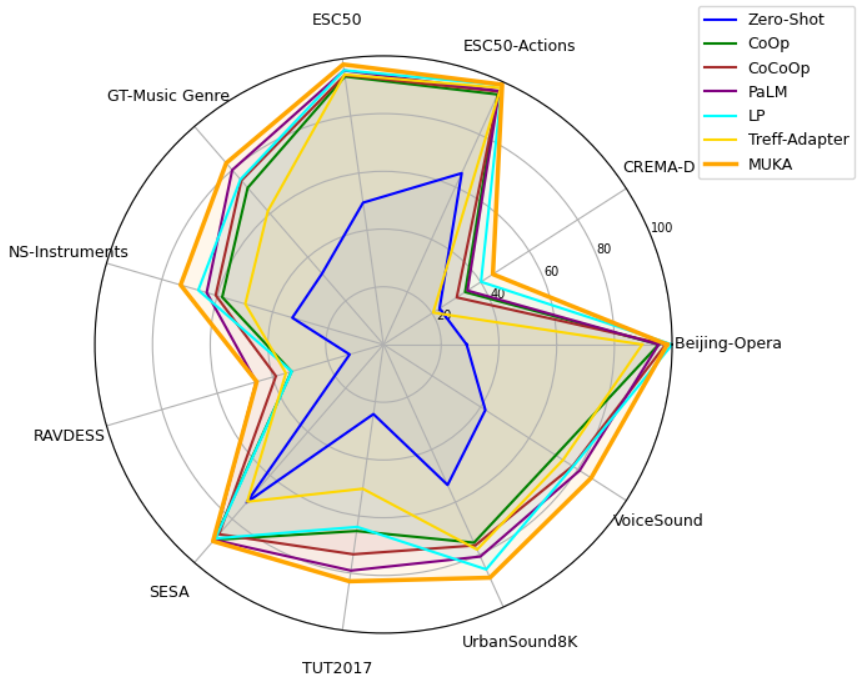}}
\caption{Comparison between few-shot adaptation methods of large audio-models.}
\label{fig:spider_chart}
\end{figure}

Recent advances in multimodal learning have demonstrated significant progress in the research community by leveraging cross-modal alignment techniques. Contrastive learning has been particularly effective in aligning text with both images and audio. CLIP (Contrastive Language-Image Pretraining)~\cite{clip} and its extensions, such as CLAP~\cite{msclap1}, learn joint embeddings by maximizing similarity between paired text-image and text-audio inputs  respectively while minimizing similarity between unpaired instances. This enables zero-shot transfer capabilities, where models can generalize to unseen concepts via natural language descriptions by computing similarities in the shared multimodal space.

Beyond contrastive approaches, autoregressive models have played crucial role in multimodal understanding. Models such as Flamingo~\cite{flamingo} and Audio Flamingo~\cite{audioflamingo} incorporate both textual and image or audio inputs for strong understanding abilities. These architectures use cross-modal attention to integrate visual and auditory cues into text-based reasoning.

Despite advancements in both contrastive and autoregressive multimodal learning, significant progress has been primarily observed in the image domain, where large-scale image-text datasets can be easily collected from the web. As a result, large models have been trained from scratch on billions of data examples~\cite{clip, openclip, dfn, siglip}. In contrast, the audio domain lacks comparably large sources of audio-text pairs necessary for training such systems from scratch. CLAP models~\cite{msclap1, laionclap, msclap2} address this limitation by leveraging pretrained audio and text encoders, fine-tuning them to align both modalities within a shared space using datasets containing 128k, 660k, and 4.6M pairs. Pengi~\cite{pengi} extends CLAP to open-ended tasks by instruction-based fine-tuning, enabling it to model longer contextual dependencies and capture fine-grained, context-dependent details. 

Given the relatively small number of available training data compared to vision models, the modality gap between audio and text encoders remains significant~\cite{drcap}. This challenge underscores the need for techniques that can better leverage multimodal encoders to improve zero-shot and few-shot learning.

In this context, while state-of-the-art (SOTA) approaches focus on training-based adaptation~\cite{palm}, we propose to take advantage of recent advances in training-free vision-language adaptation~\cite{proker} to improve both the performance and the efficiency of audio-based adapters~\cite{tref_adapter}. ProKeR~\cite{proker} formulates vision-language adaptation as a kernel ridge regression in the function space with a proximal regularization, avoiding overfitting issues observed in adapter-based methods such as Treff-Adapter~\cite{tref_adapter}.

While ProKeR focused on regularizing the solution in the Reproducing Kernel Hilbert Space (RKHS), the design of the kernel function, which captures the similarity between the training samples, has been overlooked. This is particularly important when using audio feature extractors: some, trained with instruction-tuning~\cite{pengi}, capture fine-grained details, whereas others, such as contrastive pretraining methods~\cite{msclap1}, are trained on large-scale audio–text pairs and emphasize broader acoustic and semantic representations

Building on top of ProKeR, we propose MUKA, a multi-kernel product approach that leverages the complementary nature of different pretrained encoders instead of using a single feature extractor. 

By constructing a product kernel that multiplies Pengi’s local similarity with CLAP’s global similarity, we obtain a discriminative kernel that simultaneously captures fine-grained details and context-level semantics in the audio–text space. This composition preserves the theoretical guarantees of kernel methods~\cite{bach2009exploring,duvenaud2013kernel} while enhancing ProKeR’s representational power without requiring any additional training.

Our approach MUKA outperforms both training-based and training-free baselines while remaining computationally efficient.

\section{RELATED WORK}
\label{sec:related}

\subsection{Multimodal Language Pre-trained Models} 
CLIP (Contrastive Language-Image Pre-training)~\cite{clip} emerged as a seminal model, utilizing contrastive learning to align textual and visual representations effectively, significantly enhancing image classification and retrieval tasks by leveraging natural language supervision. Extending this framework, CLAP (Contrastive Language-Audio Pretraining)~\cite{msclap1} and AudioCLIP~\cite{audioclip} further adapt the principles to audio-text modalities, enabling robust audio classification and cross-modal retrieval capabilities. These models operate by maximizing the similarity between embeddings of corresponding modalities while minimizing it for non-corresponding pairs.

Autoregressive multimodal language models have achieved significant advancements by integrating text with vision and audio modalities to enhance the understanding of visual and auditory content. Models such as Frozen~\cite{frozen} and Pengi~\cite{pengi} are trained on triplets of text-image-text or text-audio-text in a question-answering format. These models focus on training image and audio encoders and their mapping to the text decoder, while keeping the text model components frozen. Flamingo~\cite{flamingo} and Audio Flamingo~\cite{audioflamingo} utilize pre-trained and frozen image and audio encoders, focusing solely on the training of additional cross-attention layers between existing pretrained and frozen langague model layers.

\subsection{Few-shot Adaptation of Multimodal Language Models}

Few-shot adaptation has been a hot topic for vision-language models, particularly for CLIP adaptation, where methods are generally categorized into prompt learning and efficient embedding-based approaches. Prompt learning, pioneered by CoOp~\cite{coop}, learns task-specific text prompts for the language encoder~\cite{cocoop, maple}, but requires backpropagation through the encoder, resulting in slow training. To address this, efficient methods adapt in the embedding space, achieving faster training while remaining competitive. Tip-Adapter~\cite{tip_adapter} introduced a caching-based mechanism for training-free closed-form adaptation, while ProKeR~\cite{proker} revisited caching from a kernel perspective with global proximal regularization. Training-based efficient methods typically add lightweight layers such as a linear head~\cite{clip} or MLPs like CLIP-Adapter~\cite{clip_adapter}. In parallel, audio-language models have followed similar trends: Treff-Adapter~\cite{tref_adapter} extends Tip-Adapter with a cross-attention linear model, offering both training-free and fine-tuned variants, while PaLM~\cite{palm} applies prompt learning in the token embedding space. Despite these advances, audio-language adaptation remains underexplored: PaLM improves upon zero-shot classification but is computationally costly due to prompt training, whereas Treff-Adapter provides efficiency but lags behind stronger caching methods such as Tip-Adapter. In this work, we draw inspiration from recent vision-language methods and adapt them to audio-language tasks, aiming to benchmark approaches that balance performance and efficiency.

\section{METHODOLOGY}
\label{sec:methodo}
\begin{figure}[t!]
    \centering
    \includegraphics[width=0.98\linewidth]{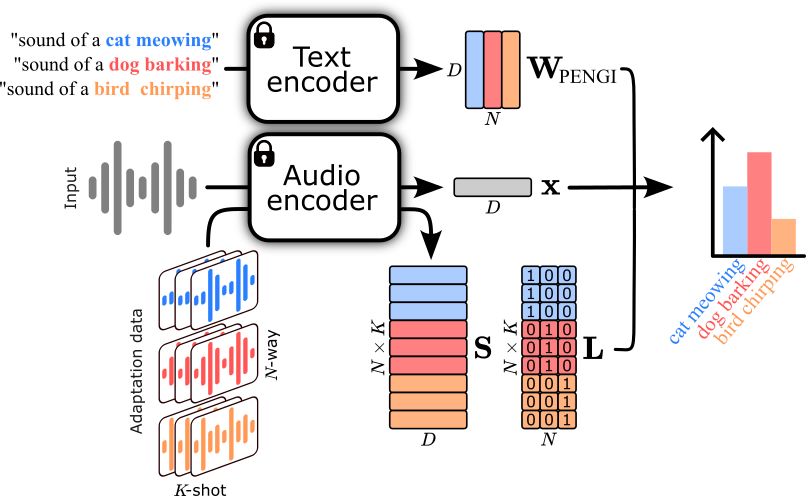}
    \caption{Training-free adaptation framework for Few-shot ALMs.}
    \label{fig:main}
\end{figure}

In this section, we formally define the different few-shot adaptation methods as shown in Figure~\ref{fig:main}. We start by defininig the classical zero-shot classification, we then provide a definition of prompt learning methods and linear probing and finally delve into cache-based methods. 
We follow the mathematical notations of ProKeR~\cite{proker}.

\subsection{Zero-shot Classification}  
The zero-shot classification leverages the text and audio encoders of PENGI to extract audio $\mathbf{x}\in\mathbb{R}^D$ and text embeddings $\mathbf{W}_{\text{PENGI}}\in\mathbb{R}^{D\times N}$, where $N$ is the number of classes. The logits of the zero-shot predictor is then defined as: 
\begin{align}
    \phi_{\text{PENGI}}(\mathbf{x}) = \mathbf{x}^{\top}\mathbf{W}_{\text{PENGI}}
\end{align}

\subsection{Training-free Adaptation Methods}  
Recent work~\cite{proker} proposed a theoretical framework for training-free few-shot VLM adaptation. In this framework, caching methods~\cite{tip_adapter, ape} are formulated as a pointwise optimal mapping defined as :
\begin{align}
    \phi(\mathbf{x}) &=~\arg\min_{\mathbf{q}\in M} \int_{M} s(\mathbf{y},\mathbf{q}) d\mu_{\mathbf{x}}(\mathbf{y}) + \mathcal{R}_{\text{clip}},
\label{eq:local-frechet}
\end{align}
where $s$ is a cost function, $\mathcal{R}_{clip}$ is a pointwise regularization using CLIP predictions (\emph{e.g.} $\mathcal{R}_{\text{clip}}~=~\lambda\|\mathbf{q}-f_{\text{clip}}(\mathbf{x})\|^2_2$  for Tip-Adapter), $d\mu_{\mathbf{x}}$ is the conditional probability of $Y$ given $X~=~\mathbf{x}$ and $M$ is the output space.

Within this framework, caching methods can be understood as the Bayes optimal solution when densities are estimated using kernel density estimators (KDE). The result is the set of non parametric kernel regression methods regularised through CLIP pointwise zero-shot predictions. 

For example, the residual Tip-Adapter~\cite{tip_adapter} is written as: 
\begin{align}
    \phi_\text{Tip}(\mathbf{x})~=~\mathbf{W}_{\text{clip}}(\mathbf{x}) + \alpha\sum\limits_{i=1}^{NK}\exp\left(-\frac{\beta}{2}\left|\left|\mathbf{S}_i-\mathbf{x}\right|\right|^2_2\right)\mathbf{L}_i,
    \label{eq:tip-adapter2}
\end{align}
where $\mathbf{W}_{\text{clip}}$ are the weights of the zero-shot CLIP classifier.

ProKeR was originally introduced for vision-language models by formulating few-shot adaptation as proximal kernel ridge regression in a reproducing kernel Hilbert space (RKHS) while conserving the benefits of training-free methods.

In essence, ProKeR is the solution of a proximal Kernel Ridge Regression problem:

\begin{multline}
    \phi(\mathbf{x})~=~\mathbf{W}_{\text{clip}} + \sum\limits_{i=1}^{NK}k(\mathbf{S_i, \mathbf{x}})\bm{\gamma}_{i}, \\ 
    \text{where} \quad \bm{\gamma}~=~(\mathbf{I}+\frac{1}{\lambda}k(\mathbf{S}, \mathbf{S}))^{-1}(\mathbf{L}-\mathbf{W}_{\text{clip}}(\mathbf{S})).
\label{equ:proker}
\end{multline}
where $k$ is the commonly used RBF kernel~\cite{proker} and $\bm{\gamma}_{i}~\in~\mathbb{R}^N$.

\newcommand{\xmark}{\ding{55}}
\begin{table*}[!t]
 \caption{Comparison of methods across 11 datasets. The table reports accuracy scores for training-based (CoOp, CoCoOp, PaLM, Linear Probing) and training-free (Zero-Shot, Treff-Adapter) methods.}
    \centering
    \setlength{\tabcolsep}{4pt}
    {%
    \begin{tabular}{l|c|c|c|c||c|c|c}
    \toprule
     & \multicolumn{4}{c||}{Training-based} & \multicolumn{3}{c}{Training-free} \\
     \midrule
    
    \multicolumn{1}{c|}{\textbf{Methods} $\rightarrow$} & \textbf{CoOp}~\cite{coop} & \textbf{CoCoOp}~\cite{cocoop} & \textbf{PaLM}~\cite{palm} & \textbf{Linear Probing} & \textbf{Zero-Shot}~\cite{pengi} & \textbf{Treff-Adapter}~\cite{tref_adapter} & \textbf{MUKA(Ours)}~\cite{proker} \\
    
    \midrule
    \textbf{Datasets} $\downarrow$ & - & - & - & - & - & - & - \\
    \midrule
    ESC50 & 93.82 & 94.27 & 95.93 & 96.00 & 49.65 & 94.48 & \textbf{98.03} \\
    Beijing-Opera & 95.34 & 97.74 & 95.33 & \textbf{100.00} & 28.81 & 89.64 & 98.30 \\
    CREMA-D & 33.62 & 30.18 & 34.59 & 40.18 & 23.10 & 20.46 & \textbf{45.06} \\
    ESC50-Actions & 95.25 & 96.34 & 96.58 & 98.83 & 65.25 & 97.75 & \textbf{99.00} \\
    GT-Music-Genre & 71.83 & 75.17 & 80.00 & 75.67 & 32.50 & 61.17 & \textbf{83.17} \\
    NS-Instruments & 58.22 & 60.58 & 63.83 & 66.85 & 32.91 & 49.89 & \textbf{73.24} \\
    RAVDESS & 33.20 & 38.83 & \textbf{45.96} & 33.12 & 12.22 & 35.23 & 45.76 \\
    SESA & 89.52 & 86.98 & 89.52 & 88.89 & 72.38 & 72.06 & \textbf{90.16} \\
    TUT2017 & 65.28 & 73.42 & 79.12 & 63.84 & 24.35 & 50.47 & \textbf{82.88} \\
    UrbanSound8K & 75.48 & 76.52 & 80.77 & 85.64 & 53.49 & 78.10 & \textbf{88.80} \\
    VocalSound & 70.96 & 77.90 & 80.78 & 78.10 & 41.97 & 73.94 & \textbf{85.52} \\
    \midrule
    \textbf{Average} & 71.14 & 73.47 & 76.58 & 75.19 & 39.69 & 65.74 & \textbf{80.90} \\
    \bottomrule
    \end{tabular}
    }
    \label{tab:main_results}
\end{table*}
\subsection{Similarity Enrichment with Multi-Kernel Product}
\label{sec:multi-kernel}

A key design element in kernel methods is the choice of the kernel function, which dictates how similarity between samples is defined. For example, using a feature space like Pengi~\cite{pengi} captures both semantic information and a wide range of fine-grained details. While helpful for detailed tasks, these features can harm classification in few-shot settings by introducing spurious correlations.

In contrast, a feature space like CLAP emphasizes coarse, global semantics but may overlook important details. To balance the strengths of both, we design a kernel function that hierarchically captures fine-grained and global semantic information. Instead of committing to a single feature space, we introduce a \emph{product kernel} that combines both representations.

Given inputs $x$ and $x'$, with embeddings $\phi_{\text{Pengi}}(x)$ and $\phi_{\text{CLAP}}(x)$, we define the similarity as:  

\begin{equation}
k(x, x') = k_{\text{Pengi}}\!\left(\phi_{\text{Pengi}}(x), \phi_{\text{Pengi}}(x')\right) \cdot 
k_{\text{CLAP}}\!\left(\phi_{\text{CLAP}}(x), \phi_{\text{CLAP}}(x')\right).
\end{equation}

This formulation leverages the fact that the product of positive semi-definite kernels is itself a valid kernel, thus preserving theoretical guarantees of kernel methods \cite{bach2009exploring,duvenaud2013kernel}. The intuition is that each encoder captures distinct but complementary aspects of the signal: Pengi provides fine-grained details, while CLAP captures general-purpose acoustic representations. By taking their product, we emphasize agreement across both feature spaces, leading to more discriminative similarity functions as this instance of  kernel composition has been shown to correspond to hierarchical similarity structures \cite{duvenaud2013kernel}.

\section{EXPERIMENTAL RESULTS}
\label{sec:results}

\subsection{Datasets}
We evaluate our approach on the same datasets as PaLM~\cite{palm}, covering diverse audio tasks: instrument classification, sound event classification, emotion recognition, vocal sound classification, surveillance sound event classification, acoustic scene classification, and music analysis.

For instrument classification, we use the Beijing-Opera dataset~\cite{beijing_opera} (four percussion instruments) and NS-Instruments~\cite{nsynth2017} (one-shot notes from ten classes).  
For sound event classification, we include ESC50~\cite{esc50} (50 environmental sounds), ESC50-Actions (10 human non-speech sounds), and UrbanSound8K~\cite{urban_sound} (10 urban noise types).  
For emotion recognition, we use CREMA-D~\cite{cremad} (six acted emotions) and RAVDESS~\cite{ravdess} (eight emotions).  
For vocal sounds, we use VocalSound~\cite{vocalsound} (six non-speech vocalizations).  
For surveillance sounds, we use SESA~\cite{sesa} (four surveillance classes).  
For acoustic scenes, we use TUT2017~\cite{tut} (15 environments).  
For music analysis, we use GT-Music-Genre~\cite{gt_music} (10 genres).

As in PaLM, we follow official splits and apply cross-validation for Beijing-Opera, ESC50, ESC50-Actions, UrbanSound8K, and TUT2017, reporting average accuracy. We also maintain the same preprocessing pipeline for reproducibility.

\subsection{Experimental Setup}  
For all experiments, we use the pre-trained PENGI model~\cite{pengi} as our Audio-Language Model (ALM), with model weights kept frozen in all adaptation methods to ensure a focus on parameter-efficient adaptation rather than full fine-tuning. Our experimental setup follows the few-shot learning protocol established in the PaLM~\cite{palm} paper, where we use 16 randomly selected samples per class from the training dataset and the entire test dataset for inference. For multi-fold datasets, we apply cross-validation and report the average accuracy across folds. Accuracy is used as the primary evaluation metric, and for all methods except Zero-Shot, we run experiments with three different random seeds and report the average performance. For Zero-Shot, we use the default text prompt template: “This is a recording of [CLASS].” We evaluate multiple SOTA adaptation methods, including prompt learning approaches such as CoOp~\cite{coop} and CoCoOp~\cite{cocoop}, where 16 context tokens are placed at the front of class names, as well as linear probing variants (standard and enhanced) and training-free caching-based methods. For CoOp, CoCoOp, and PaLM, we report results directly from the PaLM paper. All experiments are conducted on an NVIDIA RTX 3090 GPU to ensure consistency in computational resources across methods.

Motivated by the recent setting introduced by~\cite{proker}, we use ESC-50 to search for the best hypoerparameters of each method. We then transfer these hyperparameters to the 10 remaining datasets. 

We adopt the audio and text encoders from Pengi~\cite{pengi}, following the approach of PaLM~\cite{palm}. Specifically, the audio encoder is based on CLAP’s audio encoder~\cite{msclap2}, which incorporates HT-SAT~\cite{htsat}, a Swin Transformer architecture. For further details, we refer the reader to Deshmukh et al.~\cite{pengi}.

\subsection{Main Results} 

Table \ref{tab:main_results} shows the results of our experiments across 11 datasets of few-shot audio-language adaptation. Among training-free based methods, our method outperforms Treff-Adapter by a large margin and is a strong competitor to training-based methods. This highlights the flexibility of the obtained solution using a global regularization in function space through the lens of kernel methods. Regarding training-based methods, a simple yet effective linear probing approach significantly outperforms previously existing audio-language adaptation techniques~\cite{coop, cocoop, tref_adapter, palm}.  

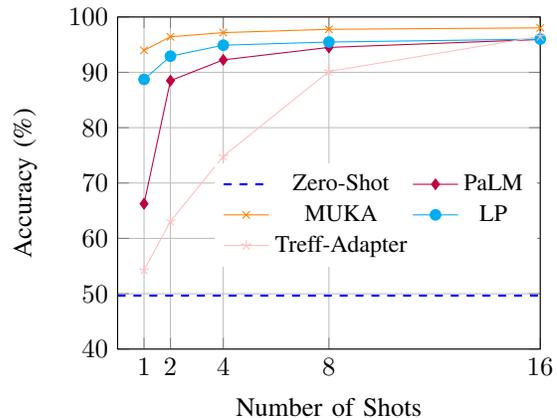
\begin{figure}[!ht]
  \centering
    \begin{tikzpicture}
        \begin{axis}[
            width=7.2cm, height=6cm,
            xlabel={Number of Shots},
            ylabel={Accuracy (\%)},
            xmin=0, xmax=16,
            ymin=40, ymax=100,
            xtick={1, 2, 4, 8, 16},
            ytick={40, 50, 60, 70, 80, 90, 100},
            legend style={
                at={(0.61,0.56)},  
                anchor=north, 
                draw=none,
                legend columns=2,
                font=\small
            },            
            grid=major,
            scaled ticks=false,
            tick label style={/pgf/number format/fixed},
            legend style={font=\small},
        ]
        
        \addplot[color=blue, mark=none, dashed, thick] coordinates {
            (0, 49.65) (16, 49.65)
        };
        \addlegendentry{Zero-Shot}

        \addplot[color=purple, mark=diamond*] coordinates {
            (1, 66.25) (2, 88.5) (4, 92.25) (8, 94.5) (16, 95.93)
        };
        \addlegendentry{PaLM}

        \addplot[color=orange, mark=x] coordinates {
            (1, 93.98) (2, 96.42) (4, 97.17) (8, 97.77) (16, 98.03)
        };
        \addlegendentry{MUKA}

        \addplot[color=cyan, mark=*] coordinates {
            (1, 88.72) (2, 92.92) (4, 94.90) (8, 95.47) (16, 96.00)
        };
        \addlegendentry{LP}

        \addplot[color=pink, mark=star] coordinates {
            (1, 54.25) (2, 63.08) (4, 74.83) (8, 90.16) (16, 96.5)
        };
        \addlegendentry{Treff-Adapter}

        \end{axis}
    \end{tikzpicture}
\caption{Few-shot accuracy on ESC-50 across different number of shots.}
\label{fig:esc50}
\end{figure}

Figure~\ref{fig:esc50} illustrates the impact of different adaptation methods on classification as the number of shots increases on ESC-50. Our method and Linear Probing exhibit rapid convergence, achieving accuracy above 97\% with as few as 4 shots, demonstrating their ability to adapt effectively with minimal data. In contrast, PaLM, though improving with additional examples, starts at a lower accuracy in 1-shot setting, indicating a greater reliance on labeled data. Meanwhile, Treff-Adapter shows the slowest progression, suggesting potential inefficiencies in leveraging small amounts of training data. These findings underscore the importance of efficient adaptation techniques for few-shot learning in audio classification.

\subsection{Ablation Study}
\begin{table}[!h]
\centering
\setlength{\tabcolsep}{6pt}
\scalebox{0.9}{
\begin{tabular}{l|c|c|c}
\textbf{} & \makecell[c]{Zero-shot \\ Predictor} & \makecell[c]{Residual \\ Feature Space} & \textbf{Avg (\%)} \\
\midrule
(a) & Pengi & Pengi         & 79.48 \\
(b) & CLAP & CLAP          & 80.29 \\
(c) & Pengi & CLAP          & 80.67 \\
(d) & Pengi & Pengi $\times$ CLAP & \textbf{80.90} \\
\end{tabular}
}
\label{tab:ablation}
\end{table}
Table~\ref{tab:ablation} summarizes our ablation results. Using Pengi exclusively (\textbf{a}), with Pengi for both zero-shot classification and kernel computation, achieves 79.48\%. Relying solely on CLAP (\textbf{b}) yields a slightly higher 80.29\%. Combining the two models brings further improvements: using Pengi for zero-shot classification while computing kernels in the CLAP space (\textbf{c}) reaches 80.67\%. Our full approach, \textbf{c}, integrates Pengi zero-shot predictions with a product of Pengi and CLAP kernels, achieving the best performance at \textbf{80.90\%}. These results highlight the complementary strengths of local detail from Pengi and global semantics from CLAP, confirming the effectiveness of multi-kernel learning.

\section{CONCLUSION}
\label{sec:conclusion}

We explored the adaptation of audio-language models across a variety of few-shot scenarios, enhancing the SOTA of both training-based and training-free methods. Our proposed approach, MUKA, leverages a multi-kernel product that combines Pengi’s fine-grained, context-dependent similarity with CLAP’s global, semantic similarity. This design not only preserves the theoretical guarantees of kernel methods but also enhances representational power without requiring additional training. By aligning local detail with global semantics in the audio–text space, MUKA achieves SOTA performance while maintaining the efficiency of training-free adaptation. In this paper, we highlight the importance of combining multiple feature spaces to design better kernel functions. This opens the venue for learnable approaches to achieve flexible data-driven kernel functions which we leave for future work.

\bibliographystyle{IEEEtran}
\bibliography{refs}

\end{document}